# Title: Magnetic Control of Valley Pseudospin in Monolayer WSe$_2$

**Authors:** G. Aivazian[1], Zhirui Gong[2], Aaron M. Jones[1], Rui-Lin Chu[3], J. Yan[4,5], D. G. Mandrus[4,5,6], Chuanwei Zhang[3], David Cobden[1], Wang Yao[2]*, X. Xu[1,7]*

**Affiliations:**

[1]Department of Physics, University of Washington, Seattle, Washington 98195, USA
[2]Department of Physics and Center of Theoretical and Computational Physics, University of Hong Kong, Hong Kong, China
[3]Department of Physics, the University of Texas at Dallas, Richardson, TX 75080 USA
[4]Materials Science and Technology Division, Oak Ridge National Laboratory, Oak Ridge, Tennessee, 37831, USA
[5]Department of Materials Science and Engineering, University of Tennessee, Knoxville, Tennessee, 37996, USA
[6]Department of Physics and Astronomy, University of Tennessee, Knoxville, Tennessee 37996, USA
[7]Department of Materials Science and Engineering, University of Washington, Seattle, Washington, 98195, USA

*Correspondence to: xuxd@uw.edu; wangyao@hku.hk

**Abstract:** Local energy extrema of the bands in momentum space, or valleys, can endow electrons in solids with pseudo-spin in addition to real spin[1-5]. In transition metal dichalcogenides this valley pseudo-spin, like real spin, is associated with a magnetic moment[1,6] which underlies the valley-dependent circular dichroism[6] that allows optical generation of valley polarization[7-9], intervalley quantum coherence[10], and the valley Hall effect[11]. However, magnetic manipulation of valley pseudospin via this magnetic moment[12-13], analogous to what is possible with real spin, has not been shown before. Here we report observation of the valley Zeeman splitting and magnetic tuning of polarization and coherence of the excitonic valley pseudospin, by performing polarization-resolved magneto-photoluminescence on monolayer WSe$_2$. Our measurements reveal both the atomic orbital and lattice contributions to the valley orbital magnetic moment; demonstrate the deviation of the band edges in the valleys from an exact massive Dirac fermion model; and reveal a striking difference between the magnetic responses of neutral and charged valley excitons which is explained by renormalization of the excitonic spectrum due to strong exchange interactions.

**Main Text**

In monolayer transition metal dichalcogenides (TMDs), there is a valley pseudospin 1/2 which describes the two inequivalent but energy degenerate band edges (the ±K valleys) at the corners of the hexagonal Brillouin zone[1]. With broken inversion symmetry, electrons in the two valleys can have finite orbital contributions to their magnetic moments which are equal in magnitude but opposite in sign by time reversal symmetry. This orbital magnetic moment is thus linked to the valley pseudospin in the same way that the bare magnetic moment ($g\mu_B S$) is linked to the real spin $S$, where $\mu_B$ is the Bohr magneton and $g$ is the Lande $g$-factor. The orbital magnetic moment in turn has two parts: a contribution from the parent atomic orbitals, and a "valley magnetic moment" contribution from the lattice structure[1] (Fig. 1a, top). The latter is related to the Berry curvature that produces the valley Hall effect[11].

The valley magnetic moment results in a valley-dependent optical selection rule in monolayer TMDs, where light of σ⁺ (σ⁻) circular polarization excites electron-hole pairs exclusively in the +K (-K) valley. This enables optical manipulation of the valley pseudospin through its excitonic states[7-10,14-17], or valley excitons. The neutral and charged valley excitons, with their exceptionally strong Coulomb interaction[18-23], are subject to a momentum-dependent gauge field arising from electron-hole exchange, or valley-orbit coupling, which at zero magnetic field is predicted to result in massless and massive dispersion respectively within the light cone[24]. This implies the possibility of controlling excitonic valley pseudospin via the Zeeman effect in an external magnetic field.

Our measurements of polarization-resolved photoluminescence (PL) in a perpendicular magnetic field are performed on mechanically exfoliated WSe$_2$ monolayers. We have obtained consistent results from many samples. The data presented here are all taken from one sample at a temperature of 30 K. In order to resolve the splitting between the +K and -K valley excitons, which is significantly smaller than the exciton linewidth (~10 meV), we both excite and detect with a single helicity of light. In this way we address one valley at a time, and the splitting can be determined by comparing the peak positions for different polarizations.

Figure 1b shows the normalized PL spectra at selected values of the magnetic field $B$. At zero field (middle) the PL from the +K valley exciton (blue, σ⁺) is identical to that from the –K valley (red, σ⁻), as expected from time-reversal symmetry[10]. In contrast, at high field the $\sigma^+$ and $\sigma^-$ components are split, with σ⁻ at a slightly higher energy than $\sigma^+$ for +7 T (top) and lower for -7 T

(bottom). We note that the small variations in lineshape seen here in the σ⁻ emission are artifacts related to sample inhomogeneity. The splitting is plotted as a function of $B$ in Fig. 1c. It is proportional to $B$ with a slope of $-0.11 \pm 0.01$ meV/T $= -(1.9 \pm 0.2)\mu_B$.

The observed magnetic spectral splitting can be explained by the combination of the magnetic moment of the tungsten $d$-orbitals[25] and the "valley magnetic moment" $m_\tau$, the lattice contribution associated with Berry curvature[6]. The bottom panel of Fig. 1a shows the Zeeman shift of the band edges from each of these two contributions as well as that from the bare spin. The dashed (solid) lines are the conduction and valence band edges at zero (positive) magnetic field, with blue and red denoting spin up and down, respectively. Because of the giant spin splitting (~ 0.4 eV) in the valence band, the valence band edge in the +K (-K) valley has only spin up (down) states. For the conduction band edge on the other hand, the spin splitting is small (~ 0.03 eV), with opposite sign in the two valleys[25-26], and both spin states are relevant.

The Zeeman shift due to the spin magnetic moment ($\Delta_s = 2s_z\mu_B B$, black arrows) does not affect the optical resonances because optical transitions conserve spin so that the effect on the initial and final states is the same. The atomic orbital contribution however does affect them because the conduction band edges are mainly composed of $d$-orbitals with $m = 0$, while the valence band edges are mainly $d$-orbitals with $m = 2$ in the +K valley and $m = -2$ in the –K valley. This contributes no shift to the conduction band and a shift of $\Delta_a = 2\tau\mu_B B$ to the valence band edge (purple arrows), resulting in a net shift of the optical resonances by $-2\tau\mu_B B$, where $\tau = \pm 1$ is the valley index for ±K.

The Zeeman shift due to the valley magnetic moment is $\Delta_v = m_\tau B$ (green arrows), with $m_\tau = \alpha_i \tau \mu_B$, where $\alpha_i$ is the valley $g$-factor for band $i$ ($i = c, v$). The leading order **k.p** approximation for the band-edge carriers yields a massive Dirac fermion model[1,6] with $\alpha = \frac{m_0}{m^*}$, where $m^*$ is the effective mass, which is the same for both conduction and valence bands[1]. Within this approximation, the valley magnetic moment therefore does not affect the exciton resonances, just as for the bare spin. However, corrections beyond leading order give different effective masses and different valley magnetic moments for the electrons and holes [25-29]. The result is a valley-dependent shift of the optical resonances by $\tau\Delta\alpha\mu_B B$, where $\Delta\alpha = \alpha_c - \alpha_v$.

The net effect is valley-dependent linear shift of the exciton resonance by $-\tau\Delta(B)/2$, where $\Delta(B) = 2(2 - \Delta\alpha)\mu_B B$ is the excitonic valley Zeeman splitting. Thus the +K valley exciton ($\tau = 1$) should be red shifted with respect to the –K one ($\tau = -1$) for $B > 0$, and blue shifted for $B < 0$, consistent with the observations in Fig. 1b. The best fit to the data (gray line) in Fig. 1c yields $\Delta\alpha = 1.1\pm0.1$. The average value of $\Delta\alpha$ found over eight samples was 0.97 (Supplementary Materials S1). This measurement of a definite non-zero $\Delta\alpha$ directly implies the existence of finite valley magnetic moments (and therefore finite Berry curvature), in addition to a deviation from the massive Dirac fermion model (Supplementary Materials S2).

The splitting in the applied magnetic field breaks the valley degeneracy, enabling control of the valley polarization. To investigate this we measure the degree of PL polarization for both helicities of incident circular polarization. Figure 2a shows PL for $\sigma^-$ excitation with $\sigma^-$ (red) and $\sigma^+$ (orange) detection at a field of -7 T. The suppression of the $\sigma^+$ signal relative to the co-polarized $\sigma^-$ peak is a signature of optically pumped valley polarization[7-10]. The degree of exciton valley polarization is clearly larger for $\sigma^+$ excitation than for $\sigma^-$ (Fig. 2b). On the other hand, when the magnetic field is reversed to +7 T (Figs. 2c and d) the polarization becomes larger for $\sigma^-$. This observation implies that, while the sign of the valley polarization is determined by the helicity of the excitation laser, its magnitude depends on the relationship between the helicity and the magnetic field direction.

Figure 2e shows the degree of PL polarization for both $\sigma^+$ (blue) and $\sigma^-$ (red) excitation as a function of $B$ between -7 T and +7 T for the neutral exciton peak. It is linear in $B$ with a negative (positive) slope. This "X" pattern implies that the valley Zeeman splitting induces an asymmetry in the intervalley scattering. (Note that the overall tilt of the "X" pattern seen here signifies an asymmetry of the response of the entire experimental system to magnetic field whose origin we do not know, but it does not affect any of our conclusions.) In contrast, the PL polarization of the negative trion peak increases for either sign of $B$ and shows a "V" pattern (Fig. 2f).

These findings can be understood as resulting from magnetic tuning of the different dispersions of valley excitons and trions, as illustrated in Fig. 3. Exchange interactions between electrons and holes strongly couple the valley pseudospin to the exciton center-of-mass wave vector **k**, splitting the exciton dispersion into two branches (Supplementary Materials S3 and S4)[24].

As shown in Fig. 3a, the upper branch has a steeper dispersion: its states have much smaller momentum compared to states at the same energy on the lower (Fig. 3a). At $B = 0$, the two branches touch at $\mathbf{k} = 0$ where the two degenerate eigenstates are the -K and +K valley excitons, which emit $\sigma^-$ and $\sigma^+$ light, respectively (Fig. 3a, middle). A finite magnetic field lifts this degeneracy and opens a gap $\Delta(B)$. For $B > 0$ the centers of the upper and lower branches are the -K and +K valley excitons respectively (Fig. 3a, right) [24]. For $B < 0$ these are interchanged (Fig. 3a, left).

The "X" pattern for neutral excitons results from the fact that $\sigma^-$ excitons form more easily for $B > 0$ and $\sigma^+$ for $B < 0$ because of the magnetic tuning of excitonic dispersion. At $B > 0$ with $\sigma^+$ excitation (Fig. 3b), electrons and holes are created at $\mathbf{k} = 0$ in the +K valley and relax to form +K excitons (blue, center of lower exciton branch) at a valley-conserving rate $\gamma_1$ or –K excitons (red, center of upper branch) at a valley-flipping rate $\gamma_2$. For $\sigma^-$ excitation (Fig. 3c), the valley-conserving and valley-flipping processes result in –K and +K valley excitons instead. The degree of PL polarization is determined by valley depolarization both during exciton formation (i.e. the ratio $\gamma_2/\gamma_1$) and in the exciton ground state before recombination ($r_1$). The measurements in Figs. 2a-d illustrate our finding that in all cases the higher energy exciton retains more valley polarization than the lower. This is opposite of what would result from thermal relaxation to the lowest energy valley exciton state. It implies that the PL polarization is largely determined during the exciton formation process.

The steeper dispersion of the upper exciton branch should facilitate formation of excitons in this branch relative to those in the lower one, in which larger momentum transfers by scattering are required to reach the light cone (c.f. supplementary figure S2). Therefore, for $B > 0$, the carriers created by $\sigma^-$ excitation have a larger valley-conserving rate $\gamma_1$ and smaller valley-flipping rate $\gamma_2$ than for $\sigma^+$, leading to $\frac{\gamma_1(\sigma^-)}{\gamma_2(\sigma^-)} > \frac{\gamma_1(\sigma^+)}{\gamma_2(\sigma^+)}$ and larger valley polarization, as observed. For $B < 0$ the converse holds by time-reversal symmetry. By solving the rate equations taking the field-dependent valley exciton formation process into account, the "X" pattern can be fully reproduced, as shown by the solid lines in Fig. 2e (Supplementary Materials S5).

Within the same framework, the "V" pattern seen for negative trions (X⁻) can also be explained as resulting from their qualitatively different spectrum. With a second electron occupying either the lowest energy spin-up band in the +K valley or spin-down band in the –K valley (top panel, Fig. 3d), at $B = 0$, X⁻ has two degenerate sets of valley-orbit coupled bands where the large exchange interaction with the extra electron opens a gap $\delta$ (~ 6 meV) [24] at $\mathbf{k} = 0$ (bottom panel, Fig. 3d). Since $\delta$ is already much larger than the achievable valley Zeeman splitting, the asymmetry in valley-exciton formation rates in the presence of a field does not dominate the $B$ dependence of the X⁻ valley polarization. Instead, the valley Zeeman splitting breaks the energy degeneracy of the X⁻ ground states, which suppresses the valley relaxation channels (grey arrows in Fig. 3e) relative to their zero-field rates. This mechanism protects the valley polarization and increases the PL polarization for either sign of $B$. Taking into account the valley depolarization in the exciton formation process, as in the neutral exciton case, we can reproduce the "V" pattern of X⁻ valley polarization and the result is the solid lines in Fig. 2f (Supplementary Materials S6).

Finally, we investigate the magnetic field dependence of valley coherence. Fig. 4a shows the polarization-resolved PL spectrum at selected fields with vertically polarized excitation and vertically (purple) and horizontally (black) polarized detection. As shown previously[10], the observed linear polarization of the exciton PL is due to the creation of valley quantum coherence. As the magnetic field increases the degree of linear polarization decreases and shows a "Λ" pattern (Fig. 4b). This demonstrates that valley coherence is suppressed by magnetic field. The data for horizontally polarized excitation is also plotted, showing that the effect is isotropic and not due to any crystal anisotropy.

Like depolarization, valley decoherence can occur during both the exciton formation and its ground state recombination processes. The magnetic field dependence in the latter case is the Hanle effect, in which spin precession quenches the linear polarization, and the half-width of the Hanle peak corresponds to the decoherence rate. The Hanle effect is qualitatively consistent with the "Λ" pattern observed; however, the extracted valley decoherence and the exciton recombination times are both on the order of 1 ps (Supplementary Materials S7), which is at least an order of magnitude smaller than those deduced from time resolved measurements[17]. It is therefore likely that valley pseudospin precession and decoherence in the exciton formation process dominates. After linearly polarized excitation generates electron-hole pairs in a superposition of the two valleys, the +K and -K valley exciton formation pathways (the red and blue wavy lines in Fig. 3b) become different as the magnetic field opens the gap. This difference destroys the optically generated coherence during the formation of ground-state excitons, leading to reduced linear polarization of the PL.

During preparation of manuscript, we became aware of similar work on $WSe_2$ by the ETH group[27] and on $MoSe_2$ by the Cornell group[28].

**Methods:** A monolayer sample of $WSe_2$ is mechanically exfoliated onto 300 nm $SiO_2$ on heavily doped Si. The samples are cooled typically to 30 K in a closed-cycle cryostat with a 7 T superconducting magnet in the Faraday geometry. The samples are excited with a 1.88 eV laser focused to a 2 μm spot size with an aspheric lens. The photoluminescence was collected with same lens and free-space coupled to a spectrometer with a liquid nitrogen cool CCD for detection.

**Acknowledgments:** We thank Xiao Li for helpful discussions. This work is mainly supported by the DoE, BES, Materials Sciences and Engineering Division (DE-SC0008145). ZG and WY were supported by the Croucher Foundation (Croucher Innovation Award), and the RGC and UGC of Hong Kong (HKU705513P, HKU9/CRF/13G, AoE/P-04/08). DC is supported by US DoE, BES, Materials Sciences and Engineering Division (DE-SC0002197). JY, DM were supported by US DoE, BES, Materials Sciences and Engineering Division. XX acknowledges a Cottrell Scholar Award. Device fabrication was performed at the University of Washington Microfabrication Facility and NSF-funded Nanotech User Facility.

**Author Contribution:** XX and WY conceived the project. GA performed the experiment, assisted by AMJ, under the supervision of XX. GA and XX analyzed the data. ZG and WY provide the theoretical explanation, with input from CL and CZ. JY and DGM synthesized and characterized the bulk $WSe_2$ crystals. GA, XX, WY, DC, and ZG wrote the paper. All authors discussed results.

**Competing Financial Interests**

The authors declare no competing financial interests.

**Figures**

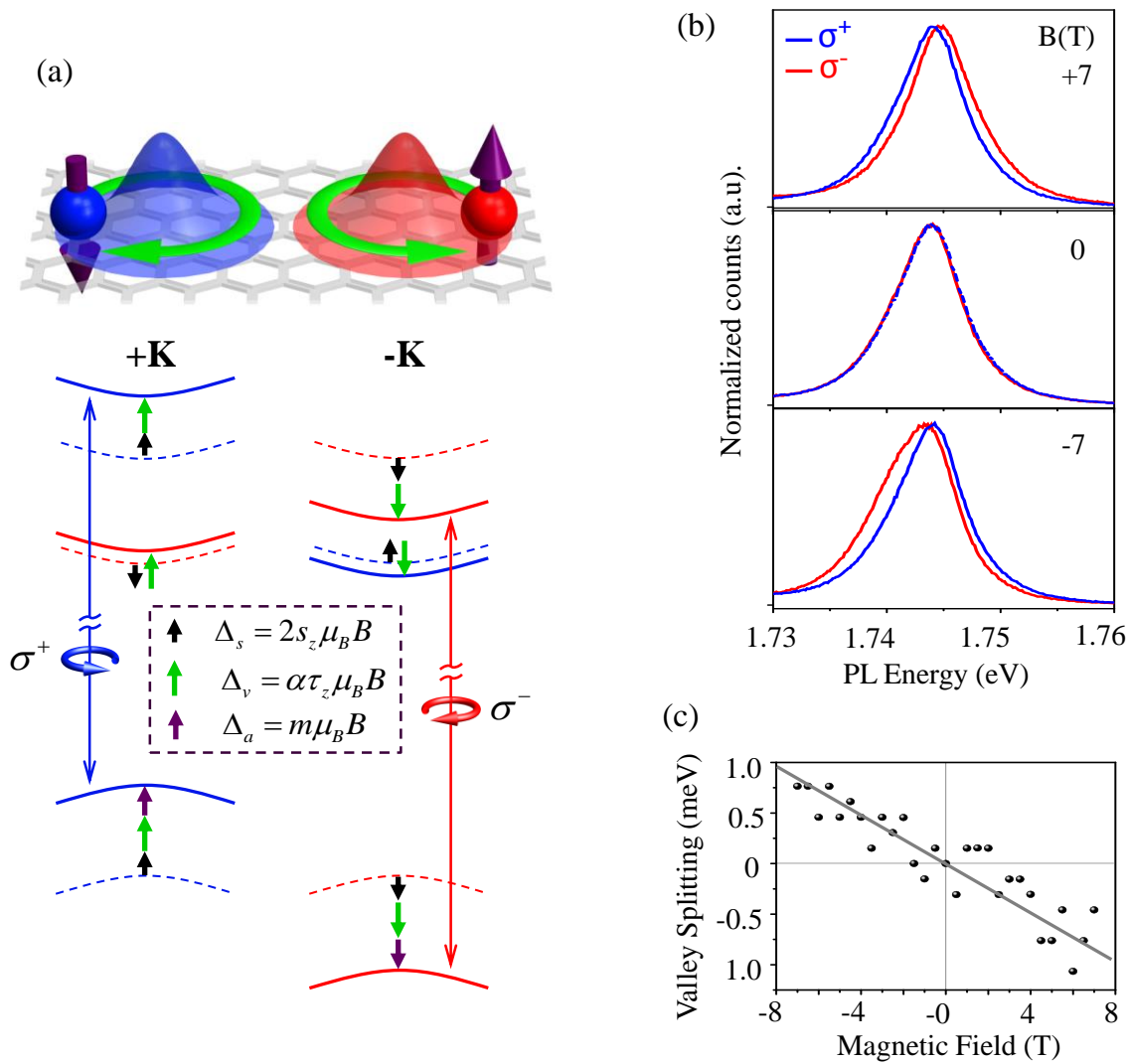

**Figure 1 | Valley Zeeman splitting. a**, Top: cartoon depicting the valley magnetic moments. Red (blue) represents spin up (down) in +K (-K) valleys. The self-rotation of the wavepacket indicated by green arrows give rise to valley magnetic moment. For holes, the magnetic moment also has a contribution from the atomic orbital (purple arrow), which has opposite sign in the K and –K valleys. Bottom: energy level diagram showing the three contributions to the valley Zeeman shifts (black for spin, green for valley, purple for atomic orbital). See text for explanation. **b**, Polarization-resolved valley exciton photoluminescence at selected magnetic fields. Blue and red curves represents photoluminescence when exciting and detecting with a single helicity, corresponding to the +K and -K valleys, respectively. **c**, Valley exciton Zeeman splitting as a function of magnetic field. The solid line is a linear fit using the equation described in the text.

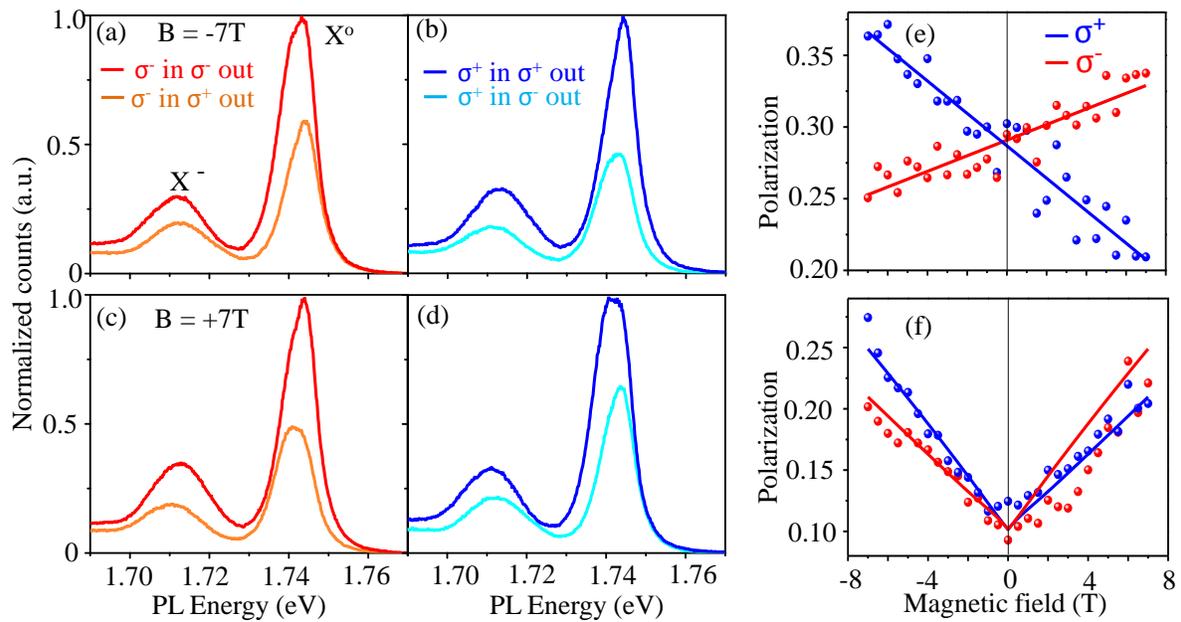

**Figure 2 | Magnetic tuning of valley polarization.** Polarization resolved photoluminescence showing the asymmetric valley pseudospin relaxation at magnetic fields of -7 T (**a** and **b**) and +7 T (**c** and **d**). **a** and **c,** for σ⁻ excitation with detection by σ⁻ (red) and σ⁺ (orange) polarization. **b** and **d,** for σ⁺ excitation and detection by σ⁺ (blue) and σ⁻ (light blue). **e** and **f**, degree of photoluminescence polarization for exciton and trion peaks, respectively. Blue (red) represents σ⁺ (σ⁻) excitation. Lines are fits using the models described in supplementary materials S5 and S6.

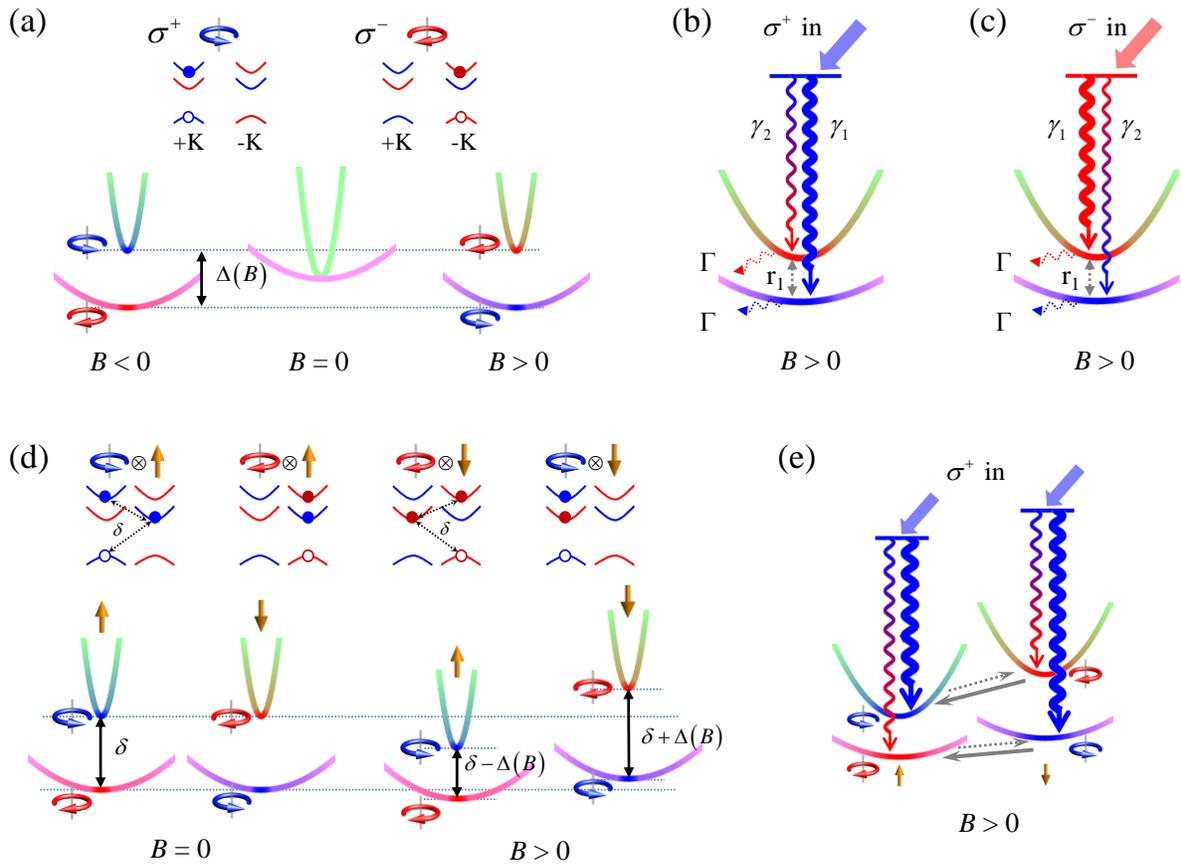

**Figure 3 | Valley exciton and trion energy spectra in a magnetic field. a**, Top: Valley excitons formed at the band edges at K and −K, which emit $\sigma^+$ and $\sigma^-$ polarized light respectively. Spin up (down) bands are shown in red (blue). Bottom: valley-orbit coupled exciton energy spectrum with and without a magnetic field. The color here indicates the valley pseudospin configuration: blue (red) denotes valley K (-K), and green and purple represent superpositions of K and −K. **b** and **c**, Cartoons depicting asymmetric valley-conserving (thick wavy lines in single color) and valley-flipping (thin wavy lines with color gradient) exciton formation process under (**b**) $\sigma^+$ and (**c**) $\sigma^-$ excitation. See text for explanation. **d**, Top: four configurations of bright trion $X^-$ labeled by valley polarization and spin orientation of the extra electron. Bottom: trion dispersion with an exchange-induced gap at **k**=0 with (right) and without (left) magnetic field. **e**, Cartoon showing asymmetric valley-conserving and valley-flipping formation of the bright trion states in a magnetic field. See text for details.

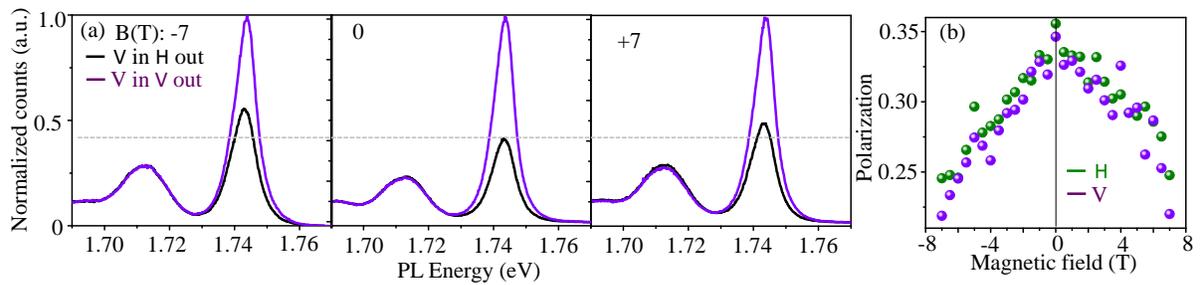

**Figure 4 | Magnetic control of valley coherence. a**, Polarization resolved photoluminescence at -7 (left), 0 (middle), and +7 T (right) by vertically polarized excitation. Purple and black curves represent vertically and horizontally polarized detection. **b**, Degree of linear polarization as a function of magnetic field. Green and purple dots denote horizontally and vertically polarized excitation.

# Supplementary Materials
# Magnetic Control of Valley Pseudospin in Monolayer WSe$_2$

Grant Aivazian, Zhirui Gong, Aaron M. Jones, Rui-Lin Chu, Jiaqiang Yan, David G. Mandrus, Chuanwei Zhang, David Cobden, Wang Yao, and Xiaodong Xu

**Supplementary Text**

**S1. Valley Zeeman effect in multiple samples.**

Eight samples were measured and all were observed to have a splitting linear in the applied field. In Figure S1 we plot the fitted slope of the splittings from all the samples, in units of Bohr magnetons. The red line is the average slope, 2.05 $\mu_B$, which corresponds to $\Delta\alpha = 0.97$. The data presented in the paper is from the last sample which is near the center of the distribution.

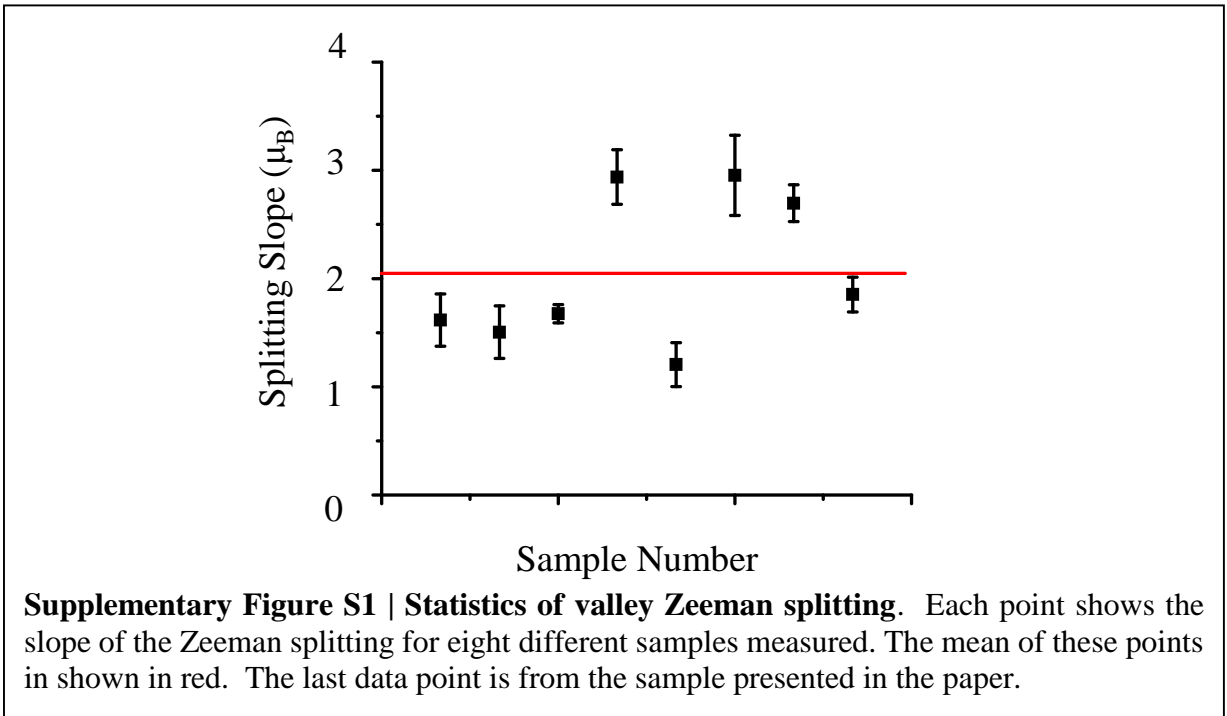

**Supplementary Figure S1 | Statistics of valley Zeeman splitting**. Each point shows the slope of the Zeeman splitting for eight different samples measured. The mean of these points in shown in red. The last data point is from the sample presented in the paper.

**S2. Effective masses and valley magnetic moment: deviation from massive Dirac model**

To the leading order approximation, the conduction and valence band edges are described by the massive Dirac fermion model and possess identical effective masses and valley magnetic moments [S1], which lead to identical Zeeman shifts of the conduction and valence bands. Thus for the Zeeman splitting of the valley exciton pseudospin, there would be no contributions from the valley magnetic moment. Nevertheless, deviation from the massive Dirac fermion model resulting from couplings to the higher energy bands will introduce a significant difference into the effective masses and valley magnetic moments of the electron and hole [S1, S2]. The difference between the valley magnetic moments of electron and hole can then contribute to the Zeeman splitting of valley exciton pseudospin.

We consider here the three-band tight-binding (TB) model with either nearest neighbor (NN) or third nearest neighbor (TNN) hoppings [S2], where the TB parameters are obtained by fitting first-principles band structures of relaxed monolayers of WSe$_2$ in both generalized-gradient approximation (GGA) and local-density approximation (LDA) cases. This is a simplest model beyond the massive Dirac fermion description of the band edges in monolayer TMDs. The results for the effective masses and the difference in the valley magnetic moments of electrons and holes are listed in Table S1. The magnetic moment is calculated with the multi-band formula in [S1]. We note that the NN and TNN models with parameters fitted from different first principle band structure calculations lead to different values on valley magnetic moments, suggesting that the models are oversimplified for quantitative description of such quantity. They are quoted here simply to illustrate the different magnetic moment of electrons and holes when one go beyond the massive Dirac fermion model.

## S3. Exciton dispersion in magnetic field and exciton formations

In monolayer TMDs, the intervalley electron-hole exchange interaction strongly couples the valley pseudospin of an exciton to its center-of-mass motion [S3]. The Hamiltonian of the valley exciton is: $H = \hbar\omega_0 + \frac{\hbar^2 k^2}{2M_0} + V'(k) + \tau_+ V_{inter}(\mathbf{k}) + \tau_- V^*_{inter}(\mathbf{k})$, where $V_{inter}(\mathbf{k}) = V(k)e^{-2i\theta}$ is from the inter-valley electron-hole exchange, $\mathbf{k} \equiv (k_x, k_y) = (k\cos\theta, k\sin\theta)$ being the center-of-mass wavevector and $\boldsymbol{\tau}$ the Pauli matrix describing the exciton valley pseudospin. The pseudospin-independent term $V'(k)$ is due to the intravalley electron-hole exchange. $M_0$ is the exciton mass, $\hbar\omega_0 \sim 1.75$ eV is the exciton energy at $\mathbf{k} = 0$, and $K$ is the distance from K to Γ point in the first Brillouin zone.

|  | NN +GGA | NN +LDA | TNN +GGA | TNN +LDA |
|---|---|---|---|---|
| $m_e^*(m_0)$ | 0.417 | 0.424 | 0.388 | 0.380 |
| $m_h^*(m_0)$ | 0.627 | 0.642 | 0.576 | 0.530 |
| $\Delta\alpha$ | 0.246 | 0.235 | 1.071 | 1.221 |

**Table S1 | Difference in the valley magnetic moments of electron and hole in monolayer WSe$_2$.** The first two rows are the effective masses of electrons and holes, and the last role lists the difference in the valley g-factor of electrons and holes, calculated with the three-band tight-binding (TB) model with either nearest neighbor (NN) or third nearest neighbor (TNN) hopping, where the TB parameters are obtained by fitting first-principles (FP) band structures of relaxed monolayers of WSe2 in both generalized-gradient approximation (GGA) and local-density approximation (LDA) cases [S2].

Because of the exceptionally strong Coulomb binding of excitons in monolayer TMDs, the electron-hole exchange is also strong, so the exciton dispersion splits into two well separated branches by the intervalley exchange. On the boundary of the light cone, the splitting between the two branches is estimated to be ~ meV. If considering the unscreened Coulomb interaction in 2D, $V(k)$ and $V'(k)$ both have linear dependence in $k$, however screening will change the dependence to quadratic. These details are not important for our discussions. The key factor here is that the upper branch has a much steeper dispersion compared to the lower branch: at the same energy, states in the lower branch correspond to much larger exciton momentum, compared to those in the upper branch, so that the exciton scatterings in the upper branch require much smaller momentum change.

At finite magnetic field applied in the perpendicular direction, the excitonic valley pseudospin spin is subject to an effective Zeeman field $-\tau_z \Delta(B)/2$, where $\Delta(B) = 2(2 - \Delta\alpha)\mu_B B$ as discussed in the main text. This Zeeman field opens a finite gap at the k=0 point, giving rise to the two-branch exciton dispersion as shown in supplementary Figure S2. In the k-space region with $V(k) \ll \Delta(B)$, the exciton eigenstates are polarized in one of the valleys (denoted by the red and blue colors respectively in Fig. S2) and coupled to circularly polarized photons. When the magnetic field changes sign, the circular polarizations of the upper and lower branches switch. In the k-space region with $V(k) \gg \Delta(B)$ excitons are linearly polarized (denoted by the green and purple colors in Fig. S2) and are not affected by the magnetic field. Only those states within the light cone can emit photons.

In the photoluminescence measurement, the excitation laser has a frequency well above $\omega_0$. Below we analyze the efficiency of bright exciton formation following the excitation by circularly polarized laser. Consider first a positive magnetic field (Figs. 3b and c, main text). The center of the upper exciton branch is then $\sigma^-$ polarized, and the center of the lower exciton branch is $\sigma^+$ polarized. Under excitation by $\sigma^-$, excitons can form at the center of the upper branch through the valley-conserving formation channel with rate $\gamma_1$, and at the center of the lower branch through the valley-flipping channel with rate $\gamma_2$. We note that the valley-flipping exciton formation process concerns only the intervalley scattering that occurs before the ground state exciton is formed. The scattering between the valley configurations of ground state exciton is modeled by a separate rate $r_1$. The valley-conserving rate is more efficient than the valley-flipping one, i.e. $\gamma_1 > \gamma_2$ as evidenced by the fact that exciton PL always has the same circular polarization with the excitation laser.

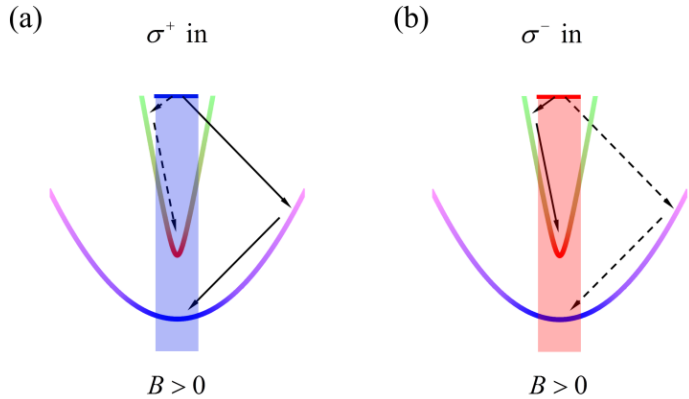

**Supplementary Figure S2 | Schematic of the exciton formation processes**. The shaded region denotes the light cone, and blue and red colors denote the (a) $\sigma^+$ or (b) $\sigma^-$ excitation respectively. The purple and green colors on the dispersion curves denote respectively linear polarization transverse or longitudinal to the momentum. For $B > 0$, comparing the valley-conserving exciton formation processes (solid arrows), the one under $\sigma^-$ excitation is facilitated in the upper exciton branch by its steeper dispersion, and is therefore more efficient than the one under $\sigma^+$ excitation which requires larger momentum transfers by scattering in the lower branch to reach the light cone. Similarly, comparing the valley-flipping exciton formations (dashed arrows), the one under $\sigma^+$ excitation is more efficient than that under $\sigma^-$ excitation.

Now we compare with the $\sigma^+$ excitation under positive magnetic field. The valley-conserving exciton formation is then at the center of the lower branch, while the valley-flipping formation is at the center of the upper branch. This can result in a difference between $\gamma_1(\sigma^+, B > 0)$ and $\gamma_1(\sigma^-, B > 0)$, denoting, respectively, the valley-conserving exciton formation rate under $\sigma^+$ and $\sigma$-excitations. This is because the upper and lower exciton branches have different dispersions. The valley-conserving exciton formation under $\sigma^-$ excitation (solid arrows, Fig. S2b) is facilitated in the upper exciton branch by its steeper dispersion, and is therefore more efficient than the valley-conserving exciton formation under $\sigma^+$ excitation (solid arrows, Fig. S2a) which requires larger momentum transfers by scattering in the lower branch to reach the light cone. Thus, we expect $\gamma_1(\sigma^+, B>0) < \gamma_1(\sigma^-, B>0)$. Similarly, comparing the valley-flipping exciton formations (dashed arrows in Fig. S2), the one under $\sigma^+$ excitation is more efficient than that under $\sigma^-$ excitation, i.e. $\gamma_2(\sigma^+, B>0) > \gamma_2(\sigma^-, B>0)$. Therefore, one may expect that $\frac{\gamma_1(\sigma^+,B>0)}{\gamma_2(\sigma^+,B>0)} < \frac{\gamma_1(B=0)}{\gamma_2(B=0)} < \frac{\gamma_1(\sigma^-,B>0)}{\gamma_2(\sigma^-,B>0)}$. The analysis is similar when the magnetic field is negative. This can qualitatively explain the X-pattern observed for the PL polarization.

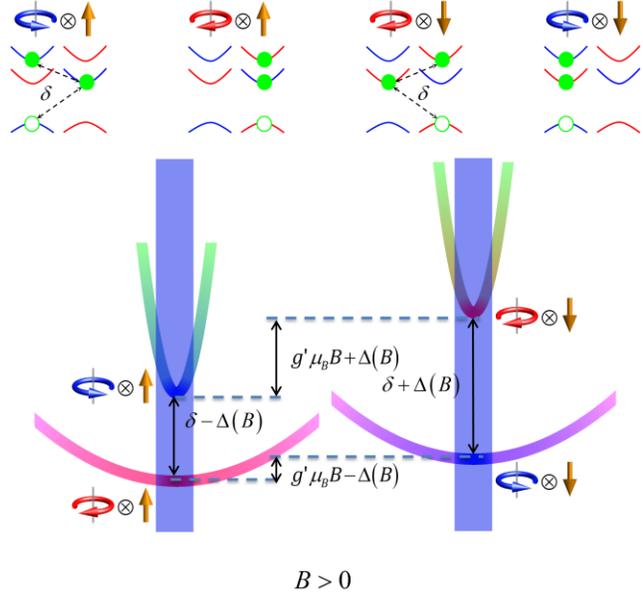

**Supplementary Figure S3:** Top: four configurations of X- labeled by the polarization of photon emission and the spin orientation of the extra electron. Bottom: X- dispersion under positive magnetic field. Left (right) is the dispersion of trion with the excess electron in the spin up state in –K (spin down state in K).

**S4. Magnetic field effects on negative trion**

Negatively charged trions (X⁻) are formed when an electron-hole pair binds an excess electron. Assuming the excess electron is in the lowest energy conduction band, X⁻ has four ground state configurations as shown in supplementary Figure S3 (top row). These four configurations can be put into two groups according to the spin configuration of the excess electron (i.e. spin down in valley K, and spin up in valley –K). The electron-hole exchange interaction couples the two configurations in each group. Hence X⁻ has two sets of valley-orbit coupled bands with the excess electron in a spin up state (at valley K) and spin down state (at –K) respectively (see Figure S3). We also need to take into account the additional exchange energy between the excess electron and the recombining electron-hole pair, which is finite for the first and third configurations shown in Figure S3 (top row), but zero for the second and fourth configurations where the spin of the excess electron is orthogonal to the other two particles. This exchange coupling is then effectively a Zeeman field in the z-direction with sign conditioned on the spin of excess electron [S3]. This opens up a gap $\delta \sim 6$ meV at zero magnetic field at k=0, where the sign of the gap depends on the spin of the excess electron (Fig. S3 bottom).

In finite magnetic field, the trion Hamiltonian is then given by

$$H_- = \hbar\omega_- + \frac{\hbar^2 k^2}{2M_-} + V'(k) + \tau_+ V_{inter}(\boldsymbol{k}) + \tau_- V_{inter}^*(\boldsymbol{k}) - \frac{\Delta(B)}{2}\tau_z - \frac{\delta}{2}\tau_z s_z + \frac{g'\mu_B B}{2} s_z \quad (1)$$

The fourth and fifth terms are the valley-orbit coupling by the intervalley electron-hole exchange, the six term is the Zeeman splitting by the magnetic field, the seventh term is the gap opened by the exchange interaction with the excess electron, and the last term is the Zeeman energy of the excess electron in the magnetic field with $g' = 2(\alpha_e - 1)$. We note that $\Delta(B) \ll \delta$ over the entire range of magnetic field in the experiment. Therefore, unlike the neutral exciton, the trion dispersions are not much affected by the magnetic field, except for the relative shift between the two sets of dispersions with the excess electron on spin up and down states respectively (see supplementary Fig. S3 bottom).

## S5. Rate equations for modeling the exciton PL polarization

Here we model the formation and valley-relaxation processes of neutral exciton and negatively charged trion with rate equations. As shown by the level scheme in Fig. 3b, under $\sigma^+$ excitation and positive magnetic field, the exciton formation and recombination processes are described by the following rate equations:

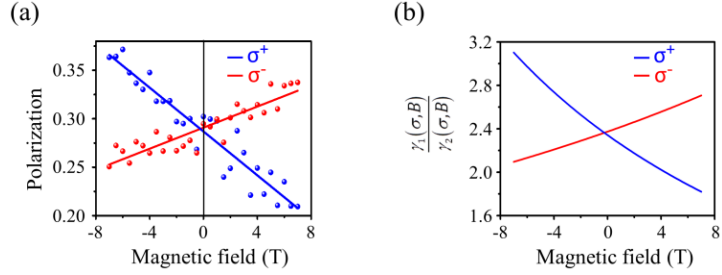

**Supplementary Figure S4:** (a) The dots are the measured polarization of the photoluminescence from exciton under $\sigma^+$ (blue) and $\sigma^-$ (red) excitation. The blue and red lines are the polarizations calculated with our model (Eq. (3)), by assuming the ratio $\gamma_1(\sigma, B)/\gamma_2(\sigma, B)$ as shown in part (b). Other parameters: $r_1 = r_1' = 0.2\Gamma$.

$$\begin{cases} \dfrac{d}{dt} N_+ = \gamma_1 - \Gamma N_+ - r_1 N_+ + r_1' N_- \\ \dfrac{d}{dt} N_- = \gamma_2 - \Gamma N_- + r_1 N_+ - r_1' N_- \end{cases}, \qquad (2)$$

where $N_\pm$ are the populations of the bright excitons in valley +K and –K within the light cone, and $\gamma_1$ and $\gamma_2$ are their corresponding formation rates (i.e. the valley-conserving and valley-flip ones, c.f. supplementary note S3). $\Gamma$ is the exciton recombination rate. $r_1'$ ($r_1$) is the relaxation rate from the higher (lower) energy valley configuration to the lower (higher) energy one in the magnetic field, and we therefore expect $r_1' \geq r_1$. So when magnetic field flips sign, $r_1'$ and $r_1$ shall be switched in Eq. (2). $\gamma_1$ and $\gamma_2$ are the valley-conserving and valley-flip exciton formation rates respectively as shown in Fig. 3b. The rate equation under $\sigma^-$ excitation is similar but with $\gamma_1$ and $\gamma_2$ switched, since the valley conserving channel now leads to formation of exciton at –K (with population $N_-$). The degree of PL polarization is given by $P_\sigma \equiv \dfrac{\sigma(N_+ - N_-)}{(N_+ + N_-)}, (\sigma = \pm 1$ for $\sigma^\pm$ excitation$)$, where positive value means the PL has the same polarization as that of the excitation.

From the steady-state solution of Eq. (2), we find

$$P_\sigma = \dfrac{\Gamma}{\Gamma + r_1' + r_1} \times \dfrac{\gamma_1 - \gamma_2}{\gamma_1 + \gamma_2} + \sigma \dfrac{B}{|B|} \dfrac{(r_1' - r_1)}{\Gamma + r_1' + r_1}. \qquad (3)$$

In the first term of the above equation, the factor $(\gamma_1 - \gamma_2)/(\gamma_1 + \gamma_2)$ corresponds to the valley depolarization in the exciton formation process. As discussed in the supplementary note S3,

$\frac{\gamma_1(\sigma^+,B>0)}{\gamma_2(\sigma^+,B>0)} < \frac{\gamma_1(B=0)}{\gamma_2(B=0)} < \frac{\gamma_1(\sigma^-,B>0)}{\gamma_2(\sigma^-,B>0)}$, and this magnetic field dependences of $\gamma_1$ and $\gamma_2$ can indeed give rise to the observed X pattern. The second term is proportional to the difference between $r_1'$ and $r_1$, which also corresponds to a X-pattern, but opposite to the one observed. This suggests that it has a small contribution in the experiment, i.e. $r_1' - r_1 \ll \Gamma$. In supplementary Figure S4, we show that with a reasonable choice of parameters, the observed X-pattern can be fitted quantitatively well. We note that if the system has time reversal symmetry in the absence of magnetic field and optical pump, then we shall expect the relation $\frac{\gamma_1(\sigma^+,B)}{\gamma_2(\sigma^+,B)} = \frac{\gamma_1(\sigma^-,-B)}{\gamma_2(\sigma^-,-B)}$, and the observed X-pattern shall be symmetric. The fact that the X pattern is asymmetric could imply that there is time reversal symmetry breaking.

### S6. Rate equations for modeling the trion PL polarization

Next we turn to the negatively charged trion ($X^-$). As illustrated by the level scheme in supplementary Figure S5, the trion formation and recombination processes are described by the following rate equations:

$$\begin{cases} \frac{d}{dt} N_+^l = \rho_\uparrow \gamma_1 - \Gamma N_+^l - r_1 N_+^l + r_1 N_-^l - r_2(\Delta_+) N_+^l + r_2'(\Delta_+) N_-^r \\ \frac{d}{dt} N_-^l = \rho_\uparrow \gamma_2 - \Gamma N_-^l + r_1 N_+^l - r_1 N_-^l - r_2(\Delta_-) N_-^l + r_2'(\Delta_-) N_+^r \\ \frac{d}{dt} N_+^r = \rho_\downarrow \gamma_1 - \Gamma N_+^r - r_1 N_+^r + r_1 N_-^r + r_2(\Delta_-) N_-^l - r_2'(\Delta_-) N_+^r \\ \frac{d}{dt} N_-^r = \rho_\downarrow \gamma_2 - \Gamma N_-^r + r_1 N_+^r - r_1 N_-^r + r_2(\Delta_+) N_+^l - r_2'(\Delta_+) N_-^r \end{cases} \quad (4)$$

where the superscript $l$ and $r$ represent the left and right set of energy dispersion with the excess electron on spin up and down state respectively (c.f. supplementary Fig. S5). Similar to the exciton case, $\gamma_1$ and $\gamma_2$ are the valley-conserving and valley-flip trion formation rates, which depends on the sign and size of the gap at k=0. However, since the exchange induced gap is much larger than the Zeeman shift in the magnetic field ($\delta \gg \Delta(B)$), the field has negligible effect on $\gamma_1$ and $\gamma_2$.

$\rho_\uparrow$ and $\rho_\downarrow$ in Eq. (4) are the portion of the spin up and spin down carriers in the electron gas in the steady state, which depends on the magnetic field B as well as the polarization $\sigma$ of the excitation light, since the circularly polarized light effectively pumps electron spin polarization. $\varsigma(\sigma, B) = \frac{\rho_\uparrow - \rho_\downarrow}{\rho_\uparrow + \rho_\downarrow}$ is then the steady-state spin polarization of the electron gas and it determines the partition of the pumping rates of the two groups of trion (i.e. with excess electron in the spin up and down states respectively).

$r_2'$ and $r_2$ here denote the relaxations between the two groups of trion as shown in Fig. 3e and supplementary Figure S5. $r_2'$ ($r_2$) is the relaxation rate from the higher (lower) energy state to the lower (higher) energy one in the magnetic field, which in general depends on the energy splitting $\Delta$.

The straightforward calculation gives the degree of PL polarization as

$$P_\sigma = V_1(\sigma, B) + V_2(\sigma, B) + X_1(\sigma, B) + X_2(\sigma, B) \quad (5)$$

where

$$V_1(\sigma, B) = P_0 \frac{\Gamma + 2r_1}{2r_1} \frac{r_1[1/R(\Delta_+) + 1/R(\Delta_-)]}{r_1[1/R(\Delta_+) + 1/R(\Delta_-)] + 1} \qquad (6)$$

$$V_2(\sigma, B) = \varsigma(\sigma, B) P_0 \frac{B}{2|B|} \frac{r(\Delta_+)/R(\Delta_+) + r(\Delta_-)/R(\Delta_-)}{r_1[1/R(\Delta_+) + 1/R(\Delta_-)] + 1}, \qquad (7)$$

$$X_1(\sigma, B) = \sigma \varsigma(\sigma, B) \frac{1}{2} \frac{\Gamma[1/R(\Delta_+) - 1/R(\Delta_-)]}{r_1[1/R(\Delta_+) + 1/R(\Delta_-)] + 1}, \qquad (8)$$

$$X_2(\sigma, B) = \sigma \frac{B}{2|B|} \frac{r(\Delta_+)/R(\Delta_+) - r(\Delta_-)/R(\Delta_-)}{r_1[1/R(\Delta_+) + 1/R(\Delta_-)] + 1} \qquad (9)$$

$\sigma = \pm 1$ for $\sigma^\pm$ excitation. In the above equations, we have defined $P_0 \equiv \frac{\Gamma}{\Gamma + 2r_1} \frac{\gamma_1 - \gamma_2}{\gamma_1 + \gamma_2}$, $R(\Delta) \equiv \Gamma + r_2'(\Delta) + r_2(\Delta)$ and $r(\Delta) \equiv r_2'(\Delta) - r_2(\Delta)$.

If the magnetic field induced splitting $\Delta_+$ and $\Delta_-$ quenches the relaxation between the two group of trions (c.f. Fig. S5), $R(\Delta)$ decreases with the increase in the magnetic field strength. Then the term $V_1(\sigma, B)$ corresponds to a V-pattern of the PL polarization, where $V_1(\sigma, B = 0) = P_0 \frac{\Gamma + 2r_1}{R(\Delta = 0) + 2r_1}$, which agrees with the main feature of the observed PL polarization in the experiments (see supplementary Figure S6). The term $X_1$ vanishes at $B = 0$, and at finite field it depends on $\varsigma(\sigma, B)$, which relies on the detail of the optical spin pumping and the spin relaxation. Nevertheless, some qualitative behaviors can be determined. First, time reversal symmetry requires $-\varsigma(\sigma, B) = \varsigma(-\sigma, -B)$. Second, for $B > 0$, optical pumping by $\sigma^-$ ($\sigma^+$) excitation tends to pump spin to the low (high) energy state in the magnetic field, so the spin polarization is higher (lower), i.e. $\varsigma(\sigma^+, B > 0) \leq \varsigma(\sigma^-, B > 0)$. Therefore $X_1$ corresponds to a X-like pattern, which can account for the difference between the V-pattern of PL polarization under $\sigma^+$ and $\sigma^-$ excitations (c.f. blue and red data points in Fig. S6 (a)).

For the other two contributions to the PL polarization, $X_2(\sigma, B)$ is also a X-like pattern, and $V_2(\sigma, B)$ is a V-like pattern. We note that $r(\Delta) \ll R(\Delta)$ is expected in the entire range of magnetic field considered. Thus $V_2(\sigma, B)$ and $X_2(\sigma, B)$ can be dropped as they are much smaller compared to $V_1(\sigma, B)$ and $X_1(\sigma, B)$ respectively.

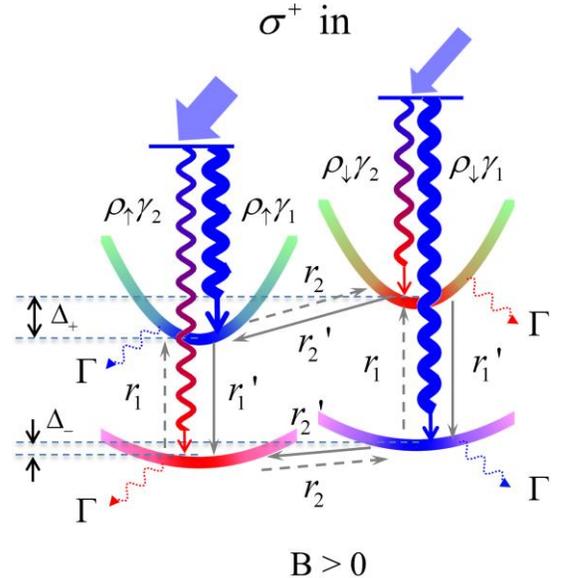

**Supplementary Figure S5: Simplified level scheme for the exciton formation and valley relaxation dynamics of trion**. See text in S6 for details.

In supplementary Figure S6, we show that with a reasonable choice of parameters, the observed V-pattern can be fitted quantitatively well.

## S7. Hanle effect

In presence of the Zeeman splitting $-\tau_z\Delta(B)/2$ along the z-direction, the in-plane valley pseudospin polarization generated by the linear polarized excitation will precess under this effective magnetic field with Larmor frequency $\Delta(B)/\hbar$. Here we lack the detail information about the timescale of the exciton formation process, so the effect of the valley pseudospin precession and decoherence during the formation process cannot be accurately counted.

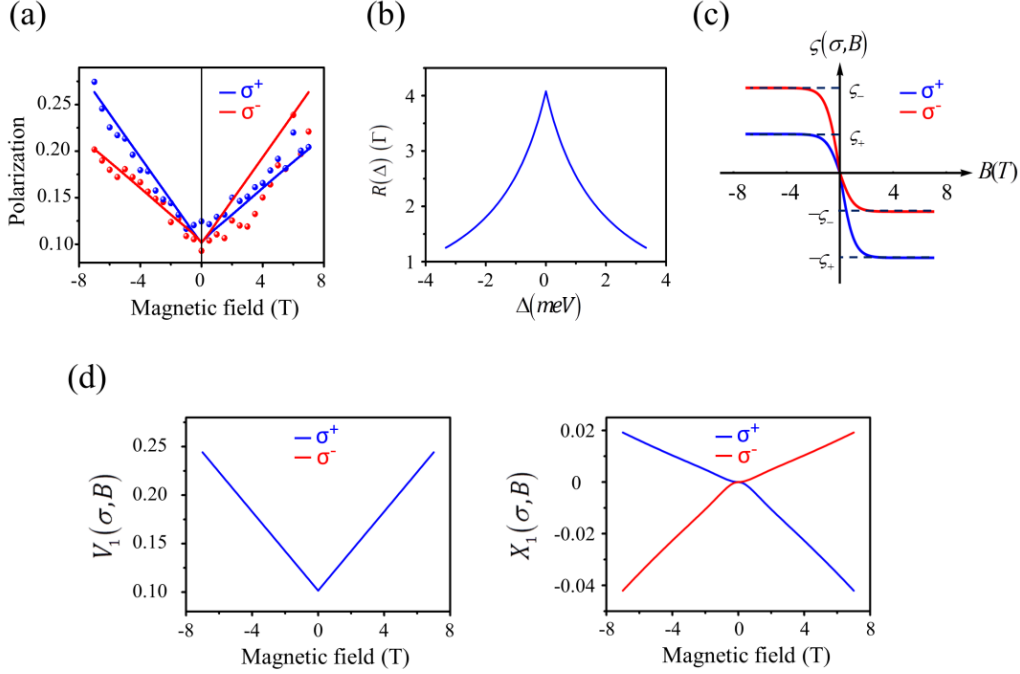

**Supplementary Figure S6:** (a) The dots are the measured polarization of photoluminescence from trion under $\sigma^+$ (blue) and $\sigma^-$ (red) excitation. The blue and red curves are the PL polarizations calculated with our model (Eq. (5-9) in supplementary note S6), by assuming $R(\Delta)$ and $\varsigma(\sigma, B)$ as shown in (b) and (c) respectively. Other parameters in the calculation: $r_1 = r_1' = 0.2\Gamma, \varsigma_+ = 0.46, \varsigma_- = 0.21, \alpha_e = 4.33$ and $\alpha_h = 3.10$. (d) The contributions from terms $V_1(\sigma, B)$ and $X_1(\sigma, B)$ respectively (c.f. Eq. (5), (6), (8) in supplementary note S6).

If we *neglect* the magnetic field effect in the exciton formation process, then the equation of motion for the valley pseudospin vector $\boldsymbol{\tau}$ of the exciton ground state is:

$$\frac{d\boldsymbol{\tau}}{dt} = -\frac{\Delta(B)}{\hbar} \times \boldsymbol{\tau} - (\Gamma^d + \Gamma)\boldsymbol{\tau} + \Gamma\boldsymbol{\tau}_0. \quad (10)$$

The first term on the right hand side describes the pseudospin precession in the effective magnetic field. In the second term, $\Gamma^d$ is the valley decoherence rate, and $\Gamma$ the exciton recombination rate. The last term describes the pumping of exciton ground state population by the linear (x) polarized excitation. $\boldsymbol{\tau}_0 = (\tau_0, 0, 0)$ corresponds to the steady-state exciton valley polarization in the limit of zero $\Gamma^d$ and $\Delta(B)$, and $\tau_0$ is determined by the exciton formation process. The steady state solution of Eq. (10) is then: $\tau_x = \tau_0 \frac{\hbar^2 \Gamma \Gamma^*}{(\hbar \Gamma^*)^2 + (\Delta(B))^2}$, $\Gamma^* \equiv \Gamma + \Gamma^d$. We note that exciton with valley pseudospin along +x (-x) direction emits photon linearly polarized in x (y). So the linear polarization of the exciton PL is $P = |\tau_x| = \frac{P_0}{1+(\Delta(B)/\hbar\Gamma^*)^2}$, $P_0 = \tau_0 \Gamma/\Gamma^*$ being the PL polarization at zero magnetic field. Because of the precession of the valley pseudospin, its time-averaged projection along the x-direction is suppressed, and hence the linear polarization of the exciton PL is quenched, which is the well-known Hanle effect [S4]. The half-width of the Hanle peak then corresponds to the decay rate $\Gamma^*$. Fitting the data in Fig. 4(b) yields $\Gamma^* \approx 1.5$ THz, and $\Gamma \approx 0.5 \tau_0^{-1}$ THz.

The above analysis *neglecting* the magnetic field effect in the exciton formation process leads to the exciton recombination lifetime and exciton valley dephasing time both of picosecond timescale, which are significantly shorter than their values obtained by time-resolved measurements [S5]. This in turn suggests that the magnetic field effect in the exciton formation process is crucial in determining the measured field dependence of linear polarization.

**Supplementary References**